\documentstyle[12pt]{article}
\begin{document}

\title{Undecidability everywhere?}
\author{K. Svozil\\
 {\small Institut f\"ur Theoretische Physik}  \\
  {\small University of Technology Vienna }     \\
  {\small Wiedner Hauptstra\ss e 8-10/136}    \\
  {\small A-1040 Vienna, Austria   }            \\
  {\small e-mail: svozil@tph.tuwien.ac.at}\\
  {\small www: http://tph.tuwien.ac.at/$\widetilde{\;\;}\,$svozil}}
\maketitle

\begin{flushright}
{\scriptsize casti.tex}
\end{flushright}

\begin{abstract}
{\em We discuss the question of if and how undecidability
might be translatable into physics, in particular with respect to
prediction and description, as well as to
complementarity games.}
\end{abstract}

\newpage

\section{Physics after the incompleteness theorems}

There is incompleteness in mathematics
\cite{godel1,turing,chaitin,calude,casti,rucker}. That means that there
does not
exist any reasonable (consistent) finite formal system from which all
mathematical truth is derivable.
There exists a ``huge'' number \cite{calude-new}
of mathematical assertions (e.g., the continuum hypothesis, the axiom
of choice) which are independent of any
particular formal system. That is, they as well as their negations are
compatible with the formal system.

Can such formal incompleteness be translated into physics or the
natural sciences in general? Is there some question about the nature of
things which is provable unknowable for rational thought?
Is it conceivable that the natural phenomena, even if they occur
deterministically, do not allow their complete description?

Before discussing these questions further, we should first clarify the
terms we use. Under a {\em physical phenomenon} we shall understand an
event, which is (irreversibly) observed. A typical physical phenomenon
consists of a click in a particle detector: there can be a click or
there can be
no click. This yes-no scheme is experimental physics in-a-nutshell (at
least according to a theoretician). From
this type of elementary observation, all of our physical
evidence is accumulated.

Then there are {\em physical theories.} Physical theories purport to
relate to the physical phenomena. At face value, they
consist of phenomena themselves: as observers, we would not be
able to know about theories if we would not observe their
representation or code. A typical code of, say, the theory
of electrodynamics, consists of the consecutive letters and symbols
printed in a book
on electrodynamics.
Their reading  corresponds to an act of observation.

Why should anyone bother about phenomena {\it versus} theories? Because
their distinction and interrelation is crucial for an understanding of
undecidability.
G\"odel, for instance, proved his incompleteness theorems as
he succeeded to properly code a (generic) theory about arithmetic {\em
within} arithmetic. (The Viennese Circle of positivists provided the
basis for the coding technique.)

At one point of the argument, we have to confront ourselves with the
question, ``is there a physical correspondent to the notion of
inconsistency; that is, to a logical contradiction \cite{hilbert-i}?''
Can a particle, for example, be here  and somewhere else
($\equiv$ not here) \cite{coh-ds}?
On the  phenomenological level, the answer is no.
To put it pointedly: there is no such thing as an inconsistent
phenomenon.
In a yes-no experiment which can have two
possible outcomes, only one of these outcomes will actually be measured.
In contradistinction, a {\em theoretical description}
might allow the consistent ``existence'' of
mutually exclusive states
if it is
indeterministic (probabilistic).
We shall come back to these issues later.

Undecidability occurs in indeterministic as well as in mechanistic
systems. The term {\em indeterminism} here stands for any process which
cannot
be described finitely and causally. As a metaphor, we may say that in
indeterministic physical systems, ``God plays dice.''
By definition, indeterminism implies undecidability. If there is no
cause, there cannot be any predictable effect. That is the whole story.
Period.

Let us be more specific and consider two examples: Firstly, in quantum
mechanics, the prevalent
probabilistic interpretation of the
wave function pretends that it is a complete description.
Single outcomes cannot be deterministically accounted for. This is the
quantum dice.
Secondly, in the scenario of ``deterministic chaos,'' the
(Martin-L\"of/Solovay/Chaitin) randomness
of ``almost all''
initial values represented by
elements of the classical
mechanical continuum is successively recovered during the time
evolution---bit after bit.
Therefore, if one believes in the quantum dice and in the physical
relevance of the classical continuum, then, by definition, there is
undecidability in physics.

Why cannot we stop here, sit back and relax?
We have just encountered the fact that present-day physical theories
contain indeterministic features which evade any complete prediction.
Why is this not the end of the story?
The trouble is that
we shall never be sure that the probabilistic interpretation of the wave
function is complete, nor do we know whether the classical continuum is
appropriate
\cite{frank}.
There might be a ``secret arena'' hidden to us momentarily,
in which everything can be deterministically accounted for.
If we relax now and uncritically accept indeterminism as a matter of
unquestionable fact, we may be heading for trouble. Indeterminism,
as it is conceived by the physics community at the {\it fin de
si\'{e}cle
(mill\'{e}naire)}, might be a degenerative research program.

Therefore, it seems not inappropriate to try to
re-interpret physical indeterminism constructively. In doing so,
it is necessary to study undecidability and incompleteness in systems
which are mechanistic in more detail. By {\em mechanistic} we mean that
they are
finitely describable and causal in all of their aspects. (In what
follows, the terms
{\em mechanistic, computable} and {\em recursive} are synonyms.) In
mechanistic
systems, every effect has a cause. But, one may doubt, if everything
has a cause, how does one
reconstruct incompleteness? The most important thing to understand at
this point is that, although in principle every effect may have a cause,
such causes might not be knowable by intrinsic observers.
We are introducing an inside-outside distinction here.
 {\em
Intrinsic observers} are embedded in the system they observe---their
``Cartesian prison''
\cite{bos,toffoli:79,roessler1,roessler2,sv:83,sv:86,sv:86a,svozil-93}.
 They cannot step outside.

The way it was defined, a mechanistic physical (dynamical) system
corresponds one-to-one to a process of computation. This computation
can, for instance be
implemented on a universal Cellular Automaton, a universal Turing
machine or any other universal computer. It, in turn, corresponds
one-to-one to
a formal system of logic. With these two identifications, namely
{\em mechanistic dynamical system}
$\equiv$
{\em computation}
$\equiv$
{\em formal system},
we bridge the gap to formal
undecidability. In principle, the term {\em system} could stand
for any of these three entities \cite{svozil-93,rasetti-95}.

We should be quite clearly aware of the fact that there is
no other possible formalization
of undecidability besides recursive function theory and formal logic. If
one resents the idea of logical or  of computer science
terminology creeping into physics, then there is no room for this
issues. Undecidability in physics marks the integration of yet another
abstract science---recursion theory---into physics.

We have set the stage now.
Let us recapitulate: we would like to consider mechanistic physical
systems.
Intrinsic observers are embedded therein. These intrinsic observers
register
physical phenomena which are operational. Moreover, they develop
theories which are intrinsically codable.
Our aim is to figure out whether or not, under these constraints,
certain
physical phenomena and theoretical propositions become undecidable.
In doing so, we have to translate the pandemonium of recursion
theoretical undecidability into physics. Our translation
guide will be the equivalence between mechanistic physical systems,
computations and formal systems.


\section{Prediction and description}
G\"odel himself did not believe in any physical relevance of the
incompleteness theorems, in particular not for quantum mechanics
\cite{bernstein}.
He might have been brainwashed
by Einstein, who was bitterly opposed to the Copenhagen interpretation
of quantum mechanics. Einstein thought that
quantum
mechanics and the Copenhagen interpretation thereof, was a
degenerative research program \cite{wheeler-Zurek:83,jammer}.
Einstein's
{\em dictum} ``God does not play dice'' has become a legend.

And yet, there is a straightforward extension of formal
incompleteness to
physics.
It is based on Turing's proof that certain propositions about
universal
computers---basically modeled to mimic elementary mechanical
paper-and-pencil operations on a sheet of paper---are undecidable.
In particular, it is impossible to predict
whether or
not a particular computation task on universal computers  will
eventually terminate
(and will output a particular result).
Therefore, if we construct a physical device capable of universal
computation,
there are some propositions about the future of this system which
are provable undecidable.

Let us be more specific and (algorithmically) prove the statement above.
It is often referred to as the ``halting theorem''  or the  ``recursive
undecidability of the halting problem.''
We are using a technique of {\em diagonalization,} which was pioneered
by Cantor in a proof of the undenumerability of the reals.
The technique is the most useful tool in exploring the undecidable.

The strategy of diagonalization (and related techniques) is to assume a
statement---whose
existence should be disproved---and, by trivial manipulations, derive a
paradox, a contradiction. The only possibility to avoid this paradox is
to abandon the original statement. For the purpose of a formal proof,
any paradox can in
principle be exploited, as long as it is codable into formal entities.
G\"odel \cite{godel1} as well as Turing \cite{turing} used ``the liar''
\cite{bible}
for their incompleteness theorems.
G\"odel was well aware of the fact that almost any
classical paradox might do as well.
(Readers not interested
in the details of the proof may skip the entire next section.)

We shall prove the recursive unsolvability of the halting problem.
Assume, for the moment,
that there is an mechanistic way to foresee a particular aspect of
the future
of an arbitrary computation. Namely, whether or not the computation
will terminate.
Or, if it outputs a string of symbols and then terminates.
Consider an arbitrary algorithm $B(x)$ whose input is a string of
symbols
$x$.
 Assume that there exists a ``predictor'' ${\tt
PREDICT}$
 which  is
 able to decide whether $B$ terminates on $x$ or not.
 Using  the predictor ${\tt PREDICT}(B(x))$, we shall construct another
  computing agent
$A$, which
 has as input any effective program $B$ and which proceeds as follows:
 Upon reading the program $B$ as input, $A$ makes a copy of it.
 This  can be readily achieved, since
 the program $B$ is presented to $A$  in some
 encoded form, i.e., as a string of symbols. In the next
 step, the agent uses the
 code of $B$ as input string for $B$ itself; i.e., $A$  forms
 $B(B)$. The agent now hands
 $B(B)$ over to the prediction subroutine
${\tt PREDICT}$.
 Then, $A$ proceeds as follows:
  if ${\tt PREDICT}(B(B))$ decides that $B(B)$
 halts, then agent
 $A$ does not halt;
this can for instance be realized by an infinite {\tt
 DO}-loop;
  if ${\tt PREDICT}(B(B))$ decides that $B(B)$
 does {\em not} halt, then
 $A$ halts. (This is the diagonalization step.)
 We shall now confront the agent $A$ with a paradoxical task by
 choosing $A$'s own code as input for itself.
Notice that $B$ is arbitrary and has not been specified yet.
The deterministic agent $A$ is representable by an algorithm with code
of $A$. Therefore, it is possible to substitute $A$ for $B$.
Assume that classically $A$ is restricted to classical bits of
information.
Then, whenever
 $A(A)$ halts,  ${\tt PREDICT}(A(A))$  forces
 $A(A)$ not to halt.
Conversely, whenever $A(A)$ does not halt, then ${\tt PREDICT}(A(A))$
 steers $A(A)$
 into the halting mode. In both cases one arrives at a
complete contradiction.
 Classically,
this contradiction can only be consistently avoided by
 assuming
 the nonexistence of $A$ and, since the only nontrivial feature of $A$
 is the use of the predictor subroutine
 ${\tt PREDICT}$, the impossibility of any such
 universal predictor.

Popper \cite{popper-48} considered these issues already in the forties.
More sophisticated  models have been put forward
by  Wolfram \cite{wolfram}, Moore
\cite{moore} and da Costa and  Doria  \cite{dacosta}.
These approaches essentially embed a universal computer
(or equivalent systems of Diophantine equations)
into a classical physical structure such as a field.
 The system is assumed
to be infinite to allow for infinite tape or its equivalent.
Then undecidability follows, for instance, from the
recursive unsolvability of the halting problem.

In short:  reasonable (consistent) theories predicting the future
behavior of arbitrary mechanistic physical systems are impossible.  So,
if one beliefs in the physical relevance of the model of universal
computers, then no physical theory can predict all the physical
phenomenology.  In particular, there are certain physical prediction
tasks which are undecidable.

But what if one insists that any computation should
remain finite? Then, in principle, it would be possible to construct a
predictor,
which would just have to simulate the system long and fast enough to
complete the prediction.
Could such a prediction take a sufficiently short time in order to be
useful? And what if a finite predictor tries to predict itself?
These questions get us into quantitative issues, which are more
involved. We shall attack them next.


The busy beaver function \cite{rado,chaitin-ACM} addresses the following
question: given a system whose  size of description is finite; more
precisely; less than or equal to
$n$ bits long. What is the biggest number $\Sigma (n)$ which can in
principle
be produced by such a system before halting (or, alternatively, before
recurring to the system's original state)?

A related question is: what is the upper bound of running time (or,
alternatively, recurrence time) of a program of length $n$ bits before
terminating?
An answer to that question gives us a feeling of how long we have to
wait for the most time-consuming program of length $n$ bits to
hold. That, of course, is a worst-case scenario. Many programs of
length $n$ bits will have halted before the maximal halting time.
Let us denote it by ${\tt TMAX}$.
We could figure out that knowledge of ${\tt TMAX}$
``solves'' the halting
problem quantitatively. Because if we knew that maximal halting
time,  then for an
arbitrary program of
$n$ bits,
we would have to wait
just a little bit longer than ${\tt TMAX}(n)$. If it would still run,
then we could be sure that it would run forever. Otherwise it would have
halted.
In this sense, knowledge of ${\tt TMAX}$ is equivalent to possessing a
perfect predictor ${\tt PREDICT}$. Since the latter one does not exist,
we expect that ${\tt TMAX}$ cannot be constructive function easy to work
with.

Indeed, Chaitin has proven
 \cite{rado,chaitin-ACM,chaitin-bb,dewdney,brady} that
 ${\tt TMAX}(n)=\Sigma (n+O(1))$ is the minimum time at which all
programs of
size smaller than or equal to $n$ bits which halt have done so.
 For large values of $n$, $\Sigma
(n)$  grows faster than any computable function
 of $n$; more precisely, let $f$
 be an arbitrary computable function, then there exists a positive
integer
$k$
 such that
$\Sigma (n)>f(n)$ for all $n>k$.

You can virtually see that any system trying to evaluate the busy beaver
function ``blows itself up.''
 Originally, T. Rado \cite{rado}
 asked how
 many $1$'s a Turing machine with $n$ possible states and an empty
 input tape
 could print on that tape before halting.
 The first values of the Turing busy beaver $\Sigma _T(x)$
 are finite and are known \cite{dewdney,brady}:
  $\Sigma _T(1)=1$,
 $\Sigma _T(2)= 4$,
  $\Sigma _T(3)=6$,
 $\Sigma _T(4)= 13$,
 $\Sigma _T(5) \ge 1915$,
 $\Sigma_T(7)\ge 22961$,
 $\Sigma_T(8)\ge 3\cdot (7\cdot 3^{92}-1)/2$.

What does all this mean for physics?
For mechanistic systems
describable by $n$ bits,
the recurrence time grows faster than any computable number
of $n$. It is therefore uncomputable and thus impossible to predict.


Any causal prediction requires a theory
 of the system which one wants to predict.
In the intrinsic observer scenario described above, there is no way to
cut out or separare the observer from the system. We have to deal with
self-description.

Can observers embedded in a system ever hope for a
complete theory or self-description?
Let us rephrase the question.
Is it possible for a system to contain a ``blueprint,'' a complete
representation, of itself?
This issue has been raised by von Neumann
in his investigation of self-reproducing automata. With such a
``blueprint'' it can be shown that  the automaton can
reconstruct a perfect replica of itself \cite{burks,rogers1,odi:89}.

To avoid confusion, it should be noted that it is never possible to have
a finite description with itself as proper part.
The trick is to employ {\em representations} or {\em names} of objects,
whose code
can be smaller than the objects themselves and can indeed be contained
in that object (cf. \cite{odi:89}, p. 165).
G\"odel's first incompleteness theorem
is such an example. Any book of electromagnetism is another.

A completely different issue is how such a theoretical self-description
can be obtained. Here, we have to make a distinction.
A complete theory or self-description might be obtained {\em passively}
from some ``intuition,''
``God'' or
``oracle.'' (Of course, one could never be sure that it is the
right one.)
But it is generally impossible for an intrinsic observer to {\em
actively}
 examine the own system and thereby
 to construct a complete theory.
One reason is self-interference and complementarity, as described
below.

As a comfort to those who conceive the ``Cartesian prison'' as the
source of all problems, one could cite a nice theorem by Gold
\cite{gold}.
It is sometimes referred to as the recursive undecidability of the
rule inference problem:
For any mechanistic intelligence/agent $A$, there exists a total
recursive function
$f$ such that $A$ does not infer $f$.
In more physical terms,
there is no systematic way of finding a deterministic law from the
(input/)output analysis of a mechanistic physical system.

An informal way to (algorithmically) prove Gold's theorem uses the
halting theorem. Suppose that it would indeed be possible to derive laws
systematically. Let us call this agent or computable function
${\tt RULE}$.
${\tt RULE}$ would have to watch the behavior of the system it analyzes.
But, any complete analysis would require the observation of
${\tt TMAX}(n)$,
where $n$ is the minimal description size of the system.
Since ${\tt TMAX}(n)$ grows faster than any computable function of $n$,
${\tt RULE}$ cannot be computable.

So even if the observer would be in a ``God-like'' position and would be
disentangled and freed from the observed system, he would still have to
cope with the situation that there is no systematic way of deriving
causal laws. It remains a rare art.


Of what use is a complete theory?
Is it
possible for an observer
in a finite amount of time to
predict the
own state completely?

An intuitive understanding of the impossibility of complete
self-comprehension  can be
obtained by considering a variant of Zeno's paradox of Achilles and the
Tortoise/Hector (called
``paradox of Tristram Shandy'' by Popper
Popper \cite{popper-48}):
In order
to predict oneself completely, one has
to predict oneself predicting oneself
completely, one has
to predict oneself predicting oneself
predicting oneself
completely, one has to $\ldots$''


\section{Complementarity Games}

The Hinduistic notion of Maya suggests
that
the world of senses is illusory, that
observations are distractive.
Plato's cage metaphor emphasizes the
distinction between
objects and
what we may be able to observe from these objects.
Some day, complementarity might be perceived as a variation of this
ancient theme.

There has been hardly any feature of quantum mechanics which has given
rise to as many fruitless speculations
as
{\em complementarity.} Intuitively, complementarity states that it is
impossible to (irreversibly) observe certain observables simultaneously
with arbitrary accuracy. The more precisely one of these
observables is measured, the less precisely can be the measurement of
other---complementary---observables. Typical examples of complementary
observables are position/momentum (velocity), angular momentum in
the x/y/z direction, and particle number/phase
\cite{peres,wheeler-Zurek:83}.

The intuition (if intuition makes any sense in the quantum domain)
behind this feature is that the act of
(irreversible) observation
of a physical system causes a loss of information by (irreversibly)
interfering with the system. Thereby, the possibility to measure other
aspects of the system is destroyed.

Well, this is not the whole story.
Indeed, there is reason to
believe that---at least up to
a certain magnitude of complexity---any measurement can be ``undone'' by
a proper reconstruction of the wave-function. A necessary condition for
this to happen is that {\em all} information about the original
measurement is lost.
Schr\"odinger, the creator of wave mechanics, liked to think of the wave
function as a sort of
{\em prediction catalog} \cite{schroedinger}. This prediction catalogue
contains
all potential information. Yet, it can be opened only at one particular
page.
The prediction catalog may be closed
before this page is read. Then it could be opened once more
at another, complementary, page.
By no way it is possible to open
the prediction catalog at one page, read and (irreversibly) memorize
(measure) the page, close it; then open it at another, complementary,
page.
(Two non-complementary pages which correspond to two co-measurable
observables can be read simultaneously.)

This may sound a little bit like Wodoo.
It is tempting to speculate that complementarity can never be modeled
by classical metaphors. Yet, classical examples abound.
A trivial one is a dark room with a ball moving in it. Suppose
that we want to measure its position and its velocity.
We first try to measure the ball's position by touch it.
This finite contact inevitably causes a finite change of the ball's
motion. Therefore,  we cannot any longer measure the initial velocity
of the ball with arbitrary position.

There are a number of more faithful classical metaphors for quantum
complementarity.
Take, for instance, Cohen's ``firefly-in-a-box'' model \cite{Coh},
Wright's urn model
\cite{Wri},
as well as Aerts' vessel model \cite{Aerts}.
In what follows,
we are going to explore a model of complementarity pioneered by Moore
\cite{e-f-moore}. It is based on extremely simple systems---probably the
simplest systems you can think of---on finite automata.
The finite automata we will consider here are objects which have a
finite number of internal states and a finite number of input and output
symbols.
Their time evolution is mechanistic and can be written down on tables in
matrix form.
 There are no build-in
infinities anywhere; no infinite tape or memory, no non-recursive bounds
on the runtime {\it et cetera.}

Let us develop {\em computational complementarity,} as it is
often called \cite{finkelstein}, as a game between you as the reader
and me as the author. The rules of the game are as follows.
I first give you all you need to know about the intrinsic workings of
the
automaton. For example, I tell you, ``if the automaton is in state 1 and
you input the symbol 2, then the automaton will make a transition into
state 2 and output the symbol 0;'' and so on.
Then I present you a black box which contains a realization of the
automaton. The black box has a keyboard, with
which you input the input symbols. It has an output display, on which
the output symbols appear. No other interfaces are allowed.
Suppose that I can choose in which initial state the automaton is at the
beginning of the game. I do not tell you this state. Your goal is to
find out by experiment which state I have chosen. You can simply guess
or relying on your luck by throwing a dice. But you can also perform
clever input-output experiments and analyze
your data in order to find out. You win if you give the correct answer.
I win if you guess incorrectly. (So, I have to be mean and select
worst-case examples).

Suppose that you try very hard. Is cleverness sufficient?
Will you always be able to uniquely determine the initial automaton
state?

The answer to that question is ``no.'' The reason for this
is that there may be situations when the input causes an irreversible
transition into a state which does not allow any
further queries about the initial state.
This is the meaning of the term
``self-interference'' mentioned above.
Any such irreversible loss of information about the initial value of the
automaton can be traced back
to many-to-one operations \cite{landauer}: different states
are mapped onto a single state with the same output. Many-to-one
operations such as ``deletion of information'' are the only
source of entropy increase
in mechanistic systems \cite{landauer,bennett}.

In the automaton case discussed above, one could, of course,
restore reversibility and recover the automaton's initial state by
Landauer's
``H\"ansel und Gretel''-strategy. That is, one could introduce an
additional marker at every many-to-one node which indicates the
previous state before the transition. But then, as the combined
automaton/marker system is reversible, going back to the initial state
erases all previous knowledge. This is analogous to the
re-opening of pages of Schr\"odinger's prediction catalog.

\begin{figure}
\begin{center}

\unitlength 1mm
\linethickness{0.4pt}
\begin{picture}(68.00,60.00)
(0,10)
\put(00.00,60.00){(a)}
\put(20.00,20.00){\circle*{2.00}}
\put(60.00,20.00){\circle*{2.00}}
\put(40.00,50.00){\circle*{2.00}}
\thicklines\put(22,19){\vector(1,0){36}}
\put(58.00,21.00){\vector(-1,0){36.00}}
\put(59.00,23.00){\vector(-2,3){16.00}}
\put(41.00,47.00){\vector(2,-3){16.00}}
\put(23.00,23.00){\vector(2,3){16.00}}
\put(37.00,47.00){\vector(-2,-3){16.00}}
\put(22.00,15.00){$1$}
\put(55.00,15.00){$2$}
\put(45.00,49.00){$3$}
\put(40.00,15.00){2,0}
\put(35.00,22.00){1,0}
\put(50.00,38.00){3,0}
\put(42.00,30.00){2,0}
\put(18.00,30.00){1,0}
\put(35.00,38.00){3,0}
\put(8.00,10.00){1,1}
\put(68.00,10.00){2,1}
\put(38.00,60.00){3,1}
\put(40.00,53.50){\circle{7.33}}
\put(36.33,54.33){\vector(0,-1){1.33}}
\put(63.00,17.67){\circle{7.33}}
\put(17.33,17.33){\circle{7.33}}
\put(66.67,16.33){\vector(0,1){1.00}}
\put(13.67,18.33){\vector(0,-1){1.00}}
\end{picture}
\\
$\;$\\
$\;$\\
\unitlength=1mm
\begin{picture}(140,60)(0,00)
\put(3.00,50.00){(b)}

\multiput(10,30)(20,0){3}{\circle*{1.5}}
\multiput(90,30)(20,0){3}{\circle*{1.5}}
\put(70,10){\circle*{1.5}}
\put(70,50){\circle*{1.5}}

\put(10,30){\line(3,1){60}}
\put(30,30){\line(2,1){40}}
\put(50,30){\line(1,1){20}}

\put(10,30){\line(3,-1){60}}
\put(30,30){\line(2,-1){40}}
\put(50,30){\line(1,-1){20}}

\put(90,30){\line(-1,1){20}}
\put(110,30){\line(-2,1){40}}
\put(130,30){\line(-3,1){60}}

\put(90,30){\line(-1,-1){20}}
\put(110,30){\line(-2,-1){40}}
\put(130,30){\line(-3,-1){60}}

\small

\put(3,29){\{1\}}
\put(23,29){\{2\}}
\put(43,29){\{3\}}

\put(92,29){\{1,2\}}
\put(112,29){\{1,3\}}
\put(132,29){\{2,3\}}

\put(69,5){$\emptyset$}
\put(64,52){\{1,2,3\}}

\end{picture}
\end{center}
\caption{\label{fig:1} (a) transition diagram of a quantum-like finite
automaton featuring
computational complementarity. Input and output symbols are separated by
a comma. Arrows indicate transitions.
(b) Hasse diagram of its propositional
structure. Lower elements imply higher ones if they are connected by
edge(s).}
\end{figure}
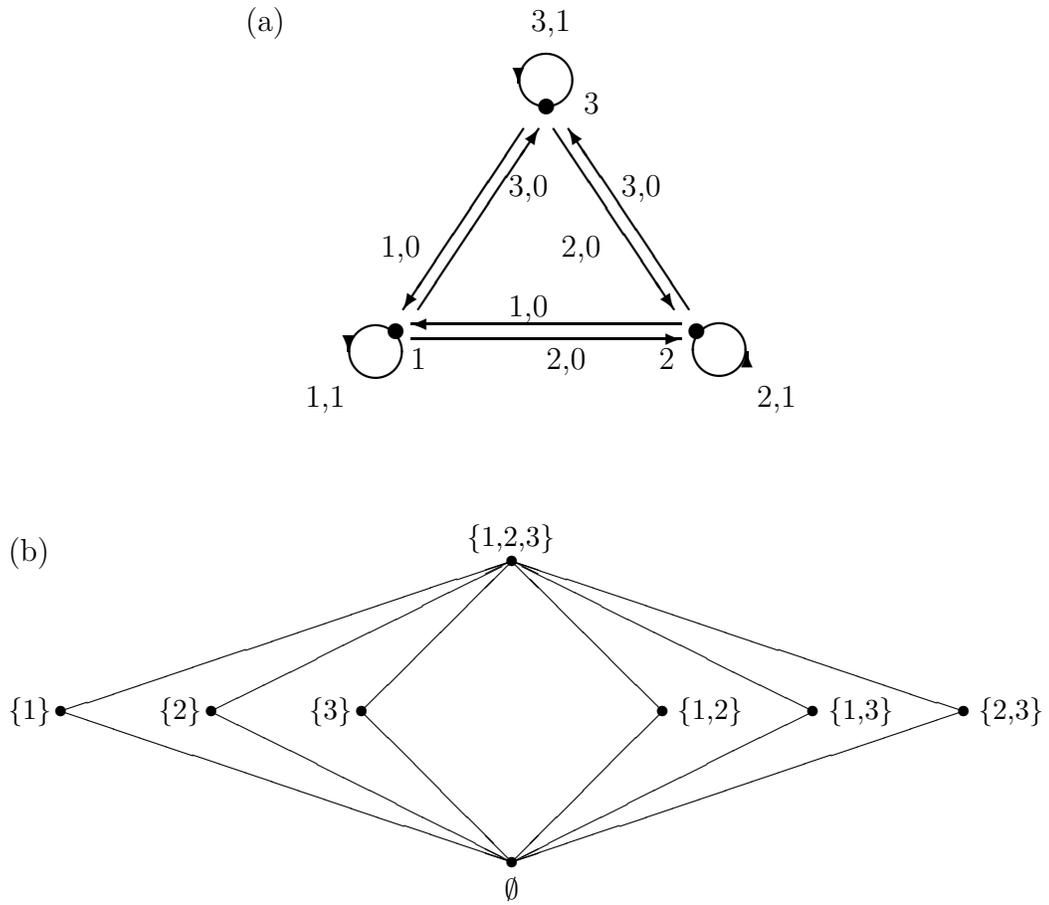

Well, this might be a good moment for introducing a sufficiently
simple example.
Consider, therefore, an automaton which can be in one of three
states, denoted by 1, 2 and 3.
This automaton accepts three input symbols, namely 1, 2 and 3.
It outputs only two symbols, namely 0 and 1.
The transition function of the automaton is as follows:
on input 1, it makes a transition to (or remains in) state 1;
on input 2, it makes a transition to (or remains in) state 2;
on input 3, it makes a transition to (or remains in) state 3.
This is a typical irreversible many-to-one operation, since a particular
input steers the automaton into that state, no matter in which one of
the three possible state it was previously.
The output function is also many-to-one and rather simple: whenever both
state and input
coincide---that is, whenever the guess was correct---it outputs 1; else
it outputs 0. So, for example, if it was in state 2 or 3 and receives
input
1, it outputs 0 and makes a transition to state 1. There it awaits
another input. These automaton specifications can be conveniently
represented by diagrams such as the one drawn in Fig.
\ref{fig:1}(a).

Computational complementarity manifests itself in the following way: if
one does not know the automaton's initial state, one has to
make choices between the input of symbols
1,2, or 3;  corresponding to definite answers whether the automaton was
in state
1, 2 or 3, corresponding to
output 1; and (2 or 3), (1 or 3) or (2 or 3), corresponding to
output 0, respectively. In the latter case, i.e., whenever the
automaton
responds with a 0 (for failure), one has lost information about the
automaton's initial state, since it surely has made a transition into
the state
1,2 or 3.
The following propositions can be stated.
On input 1, one
obtains information that the automaton either was in state 1 (exclusive)
or not in state 1, that is, in state 2 or 3. This is
denoted by $v(1)=\{ \{1\},\{2,3\}\}$.
On input 2, we
obtain information that the automaton either was in state 1 (exclusive)
or in state 1 or 3,
denoted by $v(2)=\{ \{2\},\{1,3\}\}$.
On input 3, we
obtain information that the automaton either was in state 1 (exclusive)
or in state 1 or 2,
denoted by $v(3)=\{ \{3\},\{1,2\}\}$.
In that way, we naturally arrive at the notion of a {\em partitioning}
of automaton states according to the information obtained from
input/output
experiments. Every element of the partition stands for the proposition
that the automaton is in (one of) the state(s) contained in that
partition.

{}From any partition we can construct  the
Boolean
propositional calculus which is obtained if we identify its atoms
with the elements of the partition. We then
``paste'' all Boolean propositional calculi (sometimes called
subalgebras or blocks) together.
This is a standard construction in the theory of orthomodular posets
\cite{kalmbach,ptak,piziak,navara}.
In the above example,
we arrive at a form of non-Boolean lattice whose Hasse diagram
$MO_3$ is of the ``Chinese latern'' type drawn in Fig.
\ref{fig:1}(b).

Let us go still a little bit further and ask which automaton
games of the above kind can people play.
This requires the systematic investigation of all possible
non-isomorphic automaton propositional structures, or, equivalently,
partition logics  \cite{svozil-93,schasvo1,schasvo2}. In Fig.
\ref{fig:2}, the Hasse diagrams of all nonisomorphic four-state
automaton propositional calculi are drawn.
\begin{figure}
\begin{center}
$\,$
\caption{Variations of the complementarity game up to four automaton
states. \label{fig:2}}
\end{center}
\end{figure}

Recall that the method introduced here
 is not directly
related to diagonalization and is a second, independent source of
undecidability. It is already
realizable at an elementary pre-diagonalization level,
 i.e., without
the requirement of computational universality or its arithmetic
equivalent.
The corresponding machine model is the class of finite automata.

Since any finite state
automaton can be simulated by a universal computer, complementarity is a
feature of sufficiently complex deterministic universes as well.
 To put it pointedly: if
the physical
universe is conceived as the product of a universal computation,
then complementarity is an inevitable and necessary feature of the
perception of
intrinsic    observers.
It cannot be avoided.

Conversely, any computation can be realized by a sufficiently complex
finite automaton. Therefore, the class of all complementary games is a
unique on, encompassing all possible deterministic universes.

What has all this to do with intrinsic observers? Well, during the game
one is not allowed to ``screw open'' the black box. Equivalently, one is
not allowed to make identical copies of the black box.
Both, the ``screwing open'' operation as well as copying, would
represent
actions which should only be accessible to ``God-like,'' external
observers, but not to intrinsic ones living in their ``Cartesian
prison.''
This is similar to
quantum mechanics. Copying of quantum
information (unless purely classical) or other one-to-many operations
are impossible. One cannot, for instance, produce two identical
copies of an electron or of a photon \cite{herbert}.

The complementarity game described above shows strong similarities to
quantum
mechanical systems (in two-dimensional Hilbert space).
Indeed, if we could let  the black box ``shrink'' down to point size,
we would obtain an analogue of
an electron or photon, at least for spin or polarization measurements in
three different directions. Suppose we
want to measure the spin direction of
an electron at some angle $\varphi$. We can do this by a Stern-Gerlach
device oriented in that particular direction. According to the
probabilistic interpretation of the wave function, this measurement
``randomizes''  (i.e., makes impossible any measurements of the
original) spin components in other directions. That is,
we loose information about the electron's ``original'' spin (if it is
legitimate to state that it ever had one
\cite{wheeler-Zurek:83,peres-ajp})
along the directions $\varphi' \neq \varphi$. Indeed, the propositional
structure of three spin measurements along three different angles is
identical to the one drawn in Fig. \ref{fig:1}(b).
Nevertheless, there is a difference between the ``true'' quantum
particle and its
black box-cousin: whereas the former one is supposed to have physical
spin or polarization in a {\em continuity} of directions, the latter one
can only be generalized to an arbitrary {\em countable} number of
directions. From a practical point of view, such
differences cannot be observed and are therefore operationally
irrelevant
\cite{bridgman,svozil-set}.

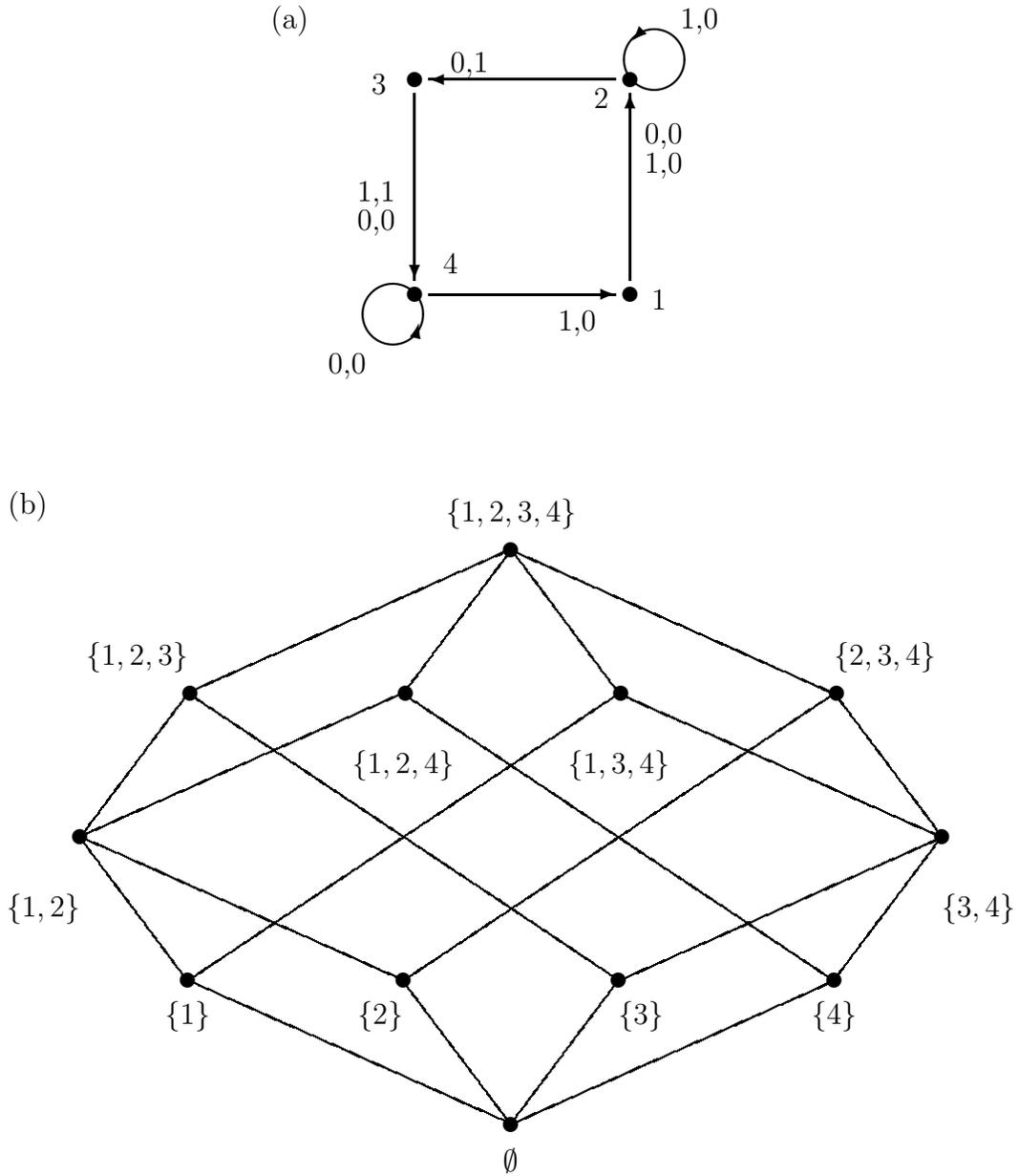
\begin{figure}
\begin{center}
\unitlength 1.00mm
\linethickness{0.4pt}
\begin{picture}(57.80,57.13)
(0,0)
\put(00.00,57.13){(a)}
\multiput(20,20)(30,0){2}{\circle*{2}}
\multiput(20,50)(30,0){2}{\circle*{2}}
\thicklines\put(22,20){\vector(1,0){26}}
\put(50.00,22.00){\vector(0,1){26.00}}
\put(48.00,50.00){\vector(-1,0){26.00}}
\put(20.00,48.00){\vector(0,-1){26.00}}
\put(53.00,18.00){$1$}
\put(45.00,46.00){$2$}
\put(14.00,48.00){$3$}
\put(24.00,23.00){$4$}
\put(52.00,37.00){1,0}
\put(52.00,41.00){0,0}
\put(25.00,51.00){0,1}
\put(12.00,29.00){0,0}
\put(12.00,33.00){1,1}
\put(40.00,15.00){1,0}
\put(57.00,57.00){1,0}
\put(8.00,9.00){0,0}
\put(53.37,52.70){\circle{9.00}}
\put(17.00,17.33){\circle{9.00}}
\put(51.00,56.00){\vector(-1,-1){0.67}}
\put(20.33,14.67){\vector(1,4){0.33}}
\end{picture}
\\
$\;$\\
$\;$\\
\unitlength 1mm
\linethickness{0.4pt}
\begin{picture}(131.00,90.00)
\put(0,90.00){(b)}
\put(55.00,25.00){\circle*{2.00}}
\put(25.00,25.00){\circle*{2.00}}
\put(85.00,25.00){\circle*{2.00}}
\put(115.00,25.00){\circle*{2.00}}
\put(55.33,65.00){\circle*{2.00}}
\put(25.33,65.00){\circle*{2.00}}
\put(85.33,65.00){\circle*{2.00}}
\put(115.33,65.00){\circle*{2.00}}
\put(70.00,5.00){\circle*{2.00}}
\put(70.00,85.00){\circle*{2.00}}
\put(130.00,45.00){\circle*{2.00}}
\put(10.00,45.00){\circle*{2.00}}
\multiput(70.00,5.00)(0.27,0.12){167}{\line(1,0){0.27}}
\multiput(115.00,25.00)(0.12,0.16){126}{\line(0,1){0.16}}
\multiput(130.00,45.00)(-0.12,0.16){126}{\line(0,1){0.16}}
\multiput(115.00,65.00)(-0.27,0.12){167}{\line(-1,0){0.27}}
\multiput(70.00,85.00)(-0.27,-0.12){167}{\line(-1,0){0.27}}
\multiput(25.00,65.00)(-0.12,-0.16){126}{\line(0,-1){0.16}}
\multiput(10.00,45.00)(0.12,-0.16){126}{\line(0,-1){0.16}}
\multiput(25.00,25.00)(0.27,-0.12){167}{\line(1,0){0.27}}
\multiput(70.00,5.00)(0.12,0.16){126}{\line(0,1){0.16}}
\multiput(85.00,25.00)(0.27,0.12){167}{\line(1,0){0.27}}
\multiput(130.00,45.00)(-0.27,0.12){167}{\line(-1,0){0.27}}
\multiput(85.00,65.00)(-0.12,0.16){126}{\line(0,1){0.16}}
\multiput(70.00,85.00)(-0.12,-0.16){126}{\line(0,-1){0.16}}
\multiput(55.00,65.00)(-0.27,-0.12){167}{\line(-1,0){0.27}}
\multiput(10.00,45.00)(0.27,-0.12){167}{\line(1,0){0.27}}
\multiput(55.00,25.00)(0.12,-0.16){126}{\line(0,-1){0.16}}
\multiput(25.00,25.00)(0.18,0.12){334}{\line(1,0){0.18}}
\multiput(115.00,25.00)(-0.18,0.12){334}{\line(-1,0){0.18}}
\multiput(25.00,65.00)(0.18,-0.12){334}{\line(1,0){0.18}}
\multiput(55.00,25.00)(0.18,0.12){334}{\line(1,0){0.18}}
\put(70.00,0.00){\makebox(0,0)[cc]{$\emptyset$}}
\put(70.00,90.00){\makebox(0,0)[cc]{$\{1,2,3,4\}$}}
\put(25.00,20.00){\makebox(0,0)[cc]{$\{1\}$}}
\put(55.00,20.00){\makebox(0,0)[rc]{$\{2\}$}}
\put(85.00,20.00){\makebox(0,0)[lc]{$\{3\}$}}
\put(115.00,20.00){\makebox(0,0)[cc]{$\{4\}$}}
\put(130.00,35.00){\makebox(0,0)[lc]{$\{3,4\}$}}
\put(10.00,35.00){\makebox(0,0)[rc]{$\{1,2\}$}}
\put(25.00,70.00){\makebox(0,0)[rc]{$\{1,2,3\}$}}
\put(55.00,55.00){\makebox(0,0)[cc]{$\{1,2,4\}$}}
\put(85.00,55.00){\makebox(0,0)[cc]{$\{1,3,4\}$}}
\put(115.00,70.00){\makebox(0,0)[lc]{$\{2,3,4\}$}}
\end{picture}
\end{center}
\caption{\label{fig:3} (a) complementarity game
featuring
weirder-than quantum properties;
(b) Hasse diagram of its propositional structure.}
\end{figure}

Even to high-ranking specialists, quantum mechanical effects appear
mindboggling \cite{green-horn-zei}.
Amazingly enough,
the complementarity game based on automata beats quantum mechanics by
weirder peculiarities.
Take, as an example, the complementarity game with the automaton drawn
in Fig.
\ref{fig:3}(a).
Input of the sequence of two symbols $00$ decides between the automaton
states 1 and 2 and 3 or 4. The resulting partition is
$v(00)=\{ \{1\},\{2\},\{3,4\}\}$.
Input of the sequence of two symbols $10$ decides between the automaton
states 1 or 2 and 3 and 4. The resulting partition is
$v(10)=\{ \{1,2\},\{3\},\{4\}\}$.
By pasting these two blocks
together, we obtain a propositional
structure represented in Fig.
\ref{fig:3}(b).

This complementarity game has several peculiar features.
 It is no lattice because the supremum and infimum is not uniquely
definable. The ``implication'' is not transitive either, because
$1\rightarrow 1\vee 2$ requires input 00 and
$1\vee 2\rightarrow 1\vee 2\vee 3$ requires input 10, whereas
$1 \rightarrow 1\vee 2\vee 3$ cannot be realized by any experiment.

It would be nice if some day the experimenters would find physical
systems
which behave in that way. Then, of course, we would have to abandon
quantum mechanics and learn some theory of the complementarity game.


\section{Should physicists really bother with undecidability?}

The program to study relative limits of knowledge can be attacked from
to opposite extreme positions. On the one hand, it may be objected that
there
are no principally unknowables, because everything is strictly causal.
On the other hand it may be stated that undecidability in physics is a
trivial matter of fact and must be accepted without any further efforts.

The first, rationalistic, position
is based on the
Cartesian
assumption
that the world is conceivable by (human) rational thought.
Laplace's demon
\cite{laplace1}
is a metaphor for this attitude.
Indeed, to many physicists, undecidability and unpredictability are
everyday
phenomena. They encounter a problem which they cannot solve or ask
questions they cannot answer.  Yet, they would hardly
conceive this experience as
an indication that there is something out there which is profoundly
undecidable.
It might not be unfair to state that---with the remarkable exceptions of
chaos and quantum theory---most physicists perceive undecidable
statements not as fundamentally unknowable but as a challenge and a
possibility for future knowledge.


Quite similarly, mathematicians often perceive G\"odel's incompleteness
theorems as artifacts.  To them, G\"odelian sentences appear
curious, even dubious,  and
explicitly constructed for their purpose.
Despite proofs that ``almost all'' true theorems are undecidable
\cite{calude-new}, they feel that all
``real''  mathematical problems they bother with
{\em are} solvable.

In this century,
the second, irrational approach, has
most influentially
been publicized---despite many reluctances from leading quantum pioneers
\cite{jammer,wheeler-Zurek:83}---by
the Copenhagen interpretation of quantum mechanics.
Chaos theory, as it is often called, has given irrationality a further
kick. Already in 1889, Poincar\'{e} suggested that certain
$n$-body problems may turn out to be impossible to solve
\cite{poincare}. Even today, after the development of recursive
(computable) function theory, many issues remain unsettled.
Certain assumptions and  problems yield the
nonpreservation of computability in classical analysis
\cite{specker,wang,kreisel,ds,pour-el,calude-sv}.
The necessity and the physical relevance of the classical continuum is
at least debatable \cite{svozil-set}.

We propose here to pursue a third path.
This third path
is characterized by
a formal investigation of the descriptive limits of theories,
as well as of predictability in general.
It may well be that this is a further, necessary step we have to go in
our understanding of Nature.

At the end, let us come back to the question posed before,
``undecidability everywere?''  It may well be that yes, there is
undecidability everywhere, and that we are confronted with it very
often.  We just may not have identified undecidability correctly, as
some emerging feature of (self-) description (with-) in a mechanistic
universe.  Depending on our philosophical assumptions, some of us may
like to think that the everday unknowns are either manifestations of
some basic randomness, a sort of ``chaos'' underlying nature; or, on the
contrary, that they are simply an artifact of our limited knowledge and
power to implement that knowledge.  We may find out that there is yet a
third possibility, having to do with with the fact that, informally
stated, self-knowledge is necessarily incomplete.

\newpage
 \tableofcontents

\begin{thebibliography}{99}
\bibitem{coh-ds}
In a double slit  experiment, the
wave function $\Psi$ is actually in a coherent superposition $\Psi
=\psi
(1)+ \psi (2)$ of the amplitude for the particle to go through slit 1
and slit 2.
\end{thebibliography}
\end{document}